\documentclass[12pt]{iopart}
\usepackage{float}
\usepackage{subfig}
\usepackage[utf8]{inputenc}
\usepackage{graphicx}
\usepackage{dcolumn}
\usepackage{bm}
\usepackage{hyperref}

\begin{document}

\title{Generalized and multiplexed $q$-plates: experimental implementation}

\author{M Vergara$^{1,2}$, E Chironi$^{1}$ and C Iemmi$^{1,2}$}

\address{$^{1}$ Facultad de Ciencias Exactas y Naturales, Departamento de Física, Universidad de Buenos Aires, Buenos Aires, Argentina.}
\address{$^{2}$ Consejo Nacional de Investigaciones Científicas y Técnicas, Buenos Aires, Argentina.}

\ead{marto@df.uba.ar}

\begin{abstract}
    In this paper we generalize the concept of $q$-plate, allowing arbitrary functions of both the radial and the azimuthal variables, and study their effect on uniformly polarized beams in the near and far-field regime. This gives a tool for achieving beams with hybrid states of polarization (SoPs), and alternative phase and intensity distributions. We also implement an experimental device based on a liquid crystal on silicon (LCoS) display for emulating these generalized $q$-plates. Moreover, we propose an application that takes advantage of the pixelated nature of this kind of devices for creating arbitrary superpositions of vector and vortex beams by representing onto the LCoS randomized combinations of two different $q$-plates, i.e. multiplexed $q$-plates. Great agreement is found between theoretical and experimental results.
\end{abstract}



\maketitle

\section{\label{sec:intro} Introduction}

Vortex beams carrying orbital angular momentum (OAM) have proven to be useful in a large number of applications ranging from classical implementations such as optical communications \cite{wang2012}, microscopy \cite{furhapter2005}, micro-manipulation \cite{bowman2011} and micro-machines design \cite{asavei2009}, to the realization of quantum information protocols in high dimensional Hilbert spaces \cite{molina2004} and multilevel quantum key distribution \cite{mirho2013}. On the other hand, vector beams, characterized by showing a non-uniform distribution of the state of polarization (SoP), have been widely studied because of their tight focusing properties \cite{quabis}, besides its potential application to communications \cite{cheng}, optical tweezers \cite{woerde}, quantum entanglement \cite{gabriel} and more.

While light propagates through a homogeneous and isotropic medium, SoP and vorticity are separately conserved; but they may be coupled in presence of anisotropic and inhomogeneous media. In 2006 Marrucci et al. introduced for this purpose the $q$-plate, which can be thought of as a half-wave retarder where the principal axis rotates with the azimuth angle \cite{marrucci2}. Hence, its matrix representation in the Jones formalism has the form 
\begin{eqnarray}
    M_q(\theta) =
    \left(\begin{array}{cc}
    \cos(2q\theta) & \sin(2q\theta) \\
    \sin(2q\theta) & -\cos(2q\theta) \end{array}\right),
\end{eqnarray}
where $2q$ is the times the principal axis of the retarder gives a whole turn around the center of the element. Although in the first years the design of $q$-plates was mainly oriented to the conversion of spin angular momentum to OAM, over time these elements evolved towards the objective of obtaining vector and vortex beams from complex superposition of SoPs and OAMs.

$Q$-plates are typically inhomogeneous and anisotropic devices where the spin to orbital conversion (STOC) is related to the Pancharatnam-Berry phase. Even though they are highly versatile elements, with many applications in the field of singular optics, different approaches extending the concept of $q$-plates were proposed in order to obtain greater flexibility in the design and diversity of responses. Some of them are based on metasurfaces \cite{Devlin896} which allow the combined use of the dynamic and geometric phases. Others, on creating $q$-plates with different $q$ values depending on the region of the element \cite{ji}, or making use of spatial light modulators (SLMs) to design $q$-plates with a nonlinear dependence of the azimuthal coordinate for binary codification \cite{holland}. In a recent paper \cite{vergara19}, we proposed the use of a generalized $q$-plate, allowing in its design arbitrary functions of the azimuthal coordinate for giving place to the generation of alternative kinds of vector and vortex beams. We also came up with a device based on a parallel aligned LCoS display, capable of achieving such distributions, emulating the generalized $q$-plate proposed.

Here we make use of this experimental device to create generalized $q$-plates with arbitrary modulations of the polarization field of a beam, in both polar coordinates $r$ and $\theta$, in such a way that we are able to explore complex vector and vortex beams, with alternative polarization and phase structures. Besides, we take the pixelated nature of the LCoS display as an advantage and propose a scheme for achieving arbitrary superposition of vector or vortex beams with different $q$ values by emulating multiplexed $q$-plates, defined as discontinuous random combinations of two different $q$-plates. Superposition of vortex beams carrying OAM has shown multiple applications, for example, for creating arbitrary OAM \textit{qudit} states for quantum information \cite{schulz13}, for optical trapping and micro-manipulation using residual OAM resulting from a superposition \cite{tao06}, or for optical communications \cite{anguita14}; while superposition of vector beams has been used for 3D polarization control \cite{li12}, improved interferometry \cite{lerman09} and more.

The Jones matrix that describes a generalized $q$-plate is
\begin{eqnarray}
    M_{\Phi}(r, \theta) =
    \left( \begin{array}{cc}
    \cos[2\Phi(r, \theta)] & \sin[2\Phi(r, \theta)] \\
    \sin[2\Phi(r, \theta)] & -\cos[2\Phi(r, \theta)] \end{array} \right),
    \label{eq:$q$-general}
\end{eqnarray}
and it represents a half-wave retarder in which the principal axis angle is an arbitrary function $\Phi(r,\theta)$.

When a linearly polarized beam passes through such an element, it becomes a vector beam with a structured linear polarization pattern in which the azimuth of the polarization vector varies as a function $2\Phi(r,\theta)$. On the other hand, when impinging with a circularly polarized beam, it acquires a phase $2\Phi(r,\theta)$, while the electric vector inverts its sense of rotation \cite{vergara19}.

This way, generalized $q$-plates allow both the generation of vortex beams with phase singularities carrying OAM, and vector beams with polarization singularities, showing many potential applications in the field of singular optics. As seen in a previous work \cite{vergara19}, interesting effects arise when these fields are propagated towards the far field regime, as the singularities tend to split giving place to different combinations.

In section \ref{sec:exp} we present the experimental device used and explain how it can manage to mimic the behaviour of a generalized $q$-plate. In section \ref{sec:rhotheta} we show simulated and experimental results for some of these generalized $q$-plates on uniformly polarized input beams, in the near and far field approximation. In section \ref{sec:multiplex} we show how to create superposition of vector or vortex beams by means of multiplexing different $q$-plates in the same element. This can be seen as a discontinuous generalized $q$-plate. Pixel by pixel modulation offered by SLMs makes this approach possible for experimental implementation. The main conclusions are given in section \ref{sec:conclus}.

\section{\label{sec:exp} Experimental device}

\begin{figure}[H]
\centering
\includegraphics[width=\columnwidth]{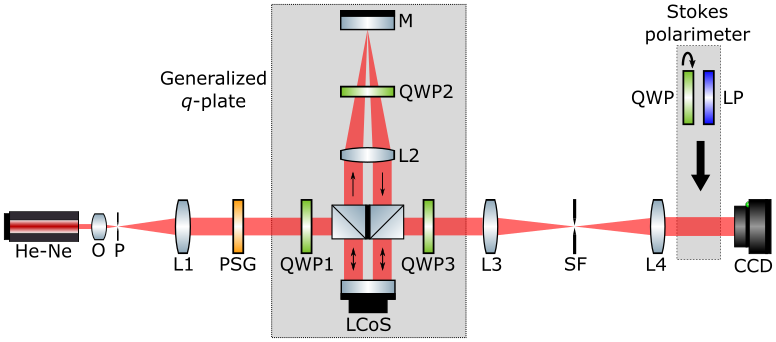}
\caption{\small Experimental device used for emulating generalized $q$-plates using a reflective PA-LCoS.}
\label{fig:dispositivo}
\end{figure}

We propose a compact device that emulates the effect of the generalized $q$-plates described above by making use of a parallel aligned reflective liquid crystal on silicon (PA-LCoS) display. In this case we use a PLUTO-NIR-010-A phase only SLM by HOLOEYE, which adds a programmable pure phase modulation to the horizontal component of the input beam. The experimental setup is sketched in Fig. \ref{fig:dispositivo}.

A He-Ne laser beam is focused on a pinhole (P) by means of a microscope objective (O), and then collimated by using a convergent lens L1. A polarization state generator (PSG) composed by a linear polarizer and a quarter wave plate is used to create arbitrary uniformly polarized beams, which then go through the generalized $q$-plate stage.

For our purpose, the first half of the SLM is programmed with a phase modulation $\zeta = -2\Phi(r,\theta) - \pi$ and the second half with a phase modulation $\eta = -\zeta = 2\Phi(r,\theta) + \pi$. The quarter wave plate QWP2 is oriented at $45^\circ$ with respect to the LC director, introducing a net $-90^\circ$ rotation of the polarization vector due to the double passage; and the lens L2, in a 4f configuration with a mirror (M) at focus, forms a real inverted image of the first half of the SLM over the second half. A pair of non-polarizing beam-splitters is used to redirect the incident beam through the 4f system (transmission of the input beam by the first beam-splitter is blocked). This way, the Jones matrix describing the whole effect of both halves of the SLM is
\begin{eqnarray}
    M_{\textnormal{SLM}}(r, \theta) =
    \left( \begin{array}{cc}
    0 & -ie^{i2\Phi(r,\theta)} \\
    ie^{-i2\Phi(r,\theta)} & 0 \end{array} \right).
    \label{eq:MSLM}
\end{eqnarray}
The phase programmed in the first half of the SLM is added to the horizontal component of the incident field which, after the 4f system, turns vertical. Then, the phase programmed in the second half is added to the former vertical component, which turns horizontal. This resembles the regular behaviour of a $q$-plate but applied to orthogonal linear states of polarization. Quarter wave plates QWP1, oriented at $45^\circ$, and QWP3, oriented at $-45^\circ$, transform the input and output beams accordingly, in order to add the respective phase modulations to circularly polarized orthogonal components of the input field, hence emulating the behaviour of a generalized $q$-plate. Jones matrix of the whole device can be calculated as
\begin{eqnarray}
\eqalign{
M &= QWP(-45^\circ)*M_{\textnormal{SLM}}(r, \theta)*QWP(45^\circ)\\
  &= \left( \begin{array}{cc}
     \cos[2\Phi(r, \theta)] & \sin[2\Phi(r, \theta)] \\
     \sin[2\Phi(r, \theta)] & -\cos[2\Phi(r, \theta)] \end{array} \right)\\
  &= M_{\Phi}(r, \theta).}
\end{eqnarray}

\begin{figure}[H]
\centering
\includegraphics[width=0.5\columnwidth]{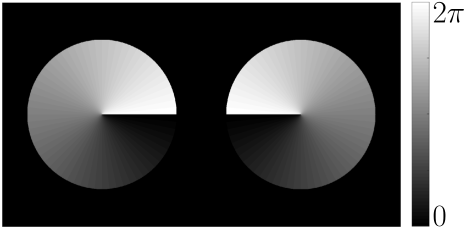}
\caption{\small Example of a phase function addressed to the SLM in order to emulate a conventional $q$-plate with $q=1/2$.}
\label{fig:slm}
\end{figure}

This matrix representation coincides with that of the generalized $q$-plate shown in Eq. \ref{eq:$q$-general}, then emulating all the expected behaviors. In addition, since it is possible to program the PA-LCoS pixel by pixel and it allows to make modifications at video rates, this scheme gives great flexibility in the design of the generalized $q$-plates. Figure \ref{fig:slm} shows an example of a phase function that can be addressed to the SLM. The function shown is the one required for emulating the conventional $q$-plate with $q=1/2$. Phase functions addressed to the SLM are flipped in order to compensate the inversion caused by lens L2 and the odd number of reflections. In the lab situation, a blazed grating is added to the phase functions in order to get the desired beam in a diffracted order on the Fourier plane, allowing to filter out the spurious light with the spatial filter SF. Also, a uniform phase value can be added to the respective phase function, in order to correct the phase shift introduced by reflection in the beam-splitters (see appendix).

Finally, the beam that emerges from the generalized $q$-plate goes through a characterization stage. Lenses L3 and L4 form a second 4f system which images the output beam onto a CCD camera. The spatial filter (SF) located at the Fourier plane is used to block out the spurious diffracted light, leaving only that coming form the programmed phase modulation. Alternatively, L4 may be replaced by a microscope objective in order to obtain a magnified image of the Fourier plane onto the CCD. A stokes polarimeter formed by a rotating quarter wave plate (QWP) and a fixed linear polarizer (LP) is used to perform polarization measurements.

\section{\label{sec:rhotheta}Generalized $q$-plates with radial and azimuthal dependence}

Beams created from $q$-plates with arbitrary functions $\Phi$ of the azimuthal and radial coordinates may show a variety of interesting effects in their amplitude, phase and polarization structure. In this section we show some examples of the different behaviors found.

\subsection{Polynomial growth in $\theta$}

First we study functions that depend only on the azimuthal coordinate, including regular $q$-plates. The first example is the non-linear function $\Phi(\theta) = q(2\pi)^{(1-p)}\theta^p$, with $p$ any integer power. The multiplicative constant $(2\pi)^{(1-p)}$ fixes the total azimuthal variation in $q$ times $2\pi$, avoiding discontinuous phase steps after a $2\pi$ period in $\theta$. These beams illustrate in a simple way the effect of non-linearities in the q-plate function.

\begin{figure}[H]
\centering
\includegraphics[width=0.5\columnwidth]{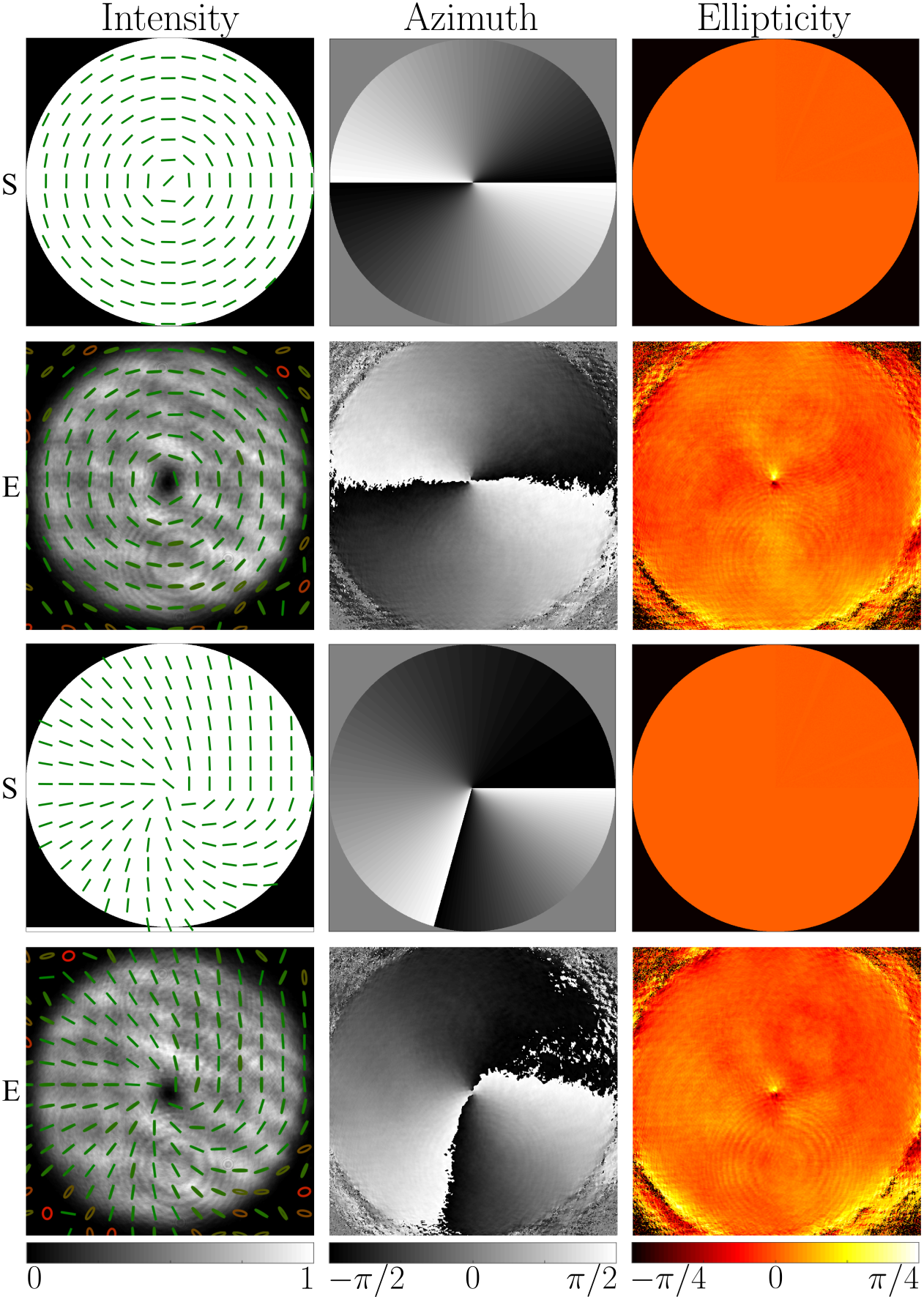}
\caption{\small Near field results for polynomial generalized q-plates with $q=1/2$, $p=1$ (first and second row) and $p=2$ (third and fourth row), when light emerging from the PSG is vertically polarized. S and E stand for simulation and experiment, respectively.}
\label{fig:polthetaV_near}
\end{figure}

\begin{figure}[H]
\centering
\includegraphics[width=0.5\columnwidth]{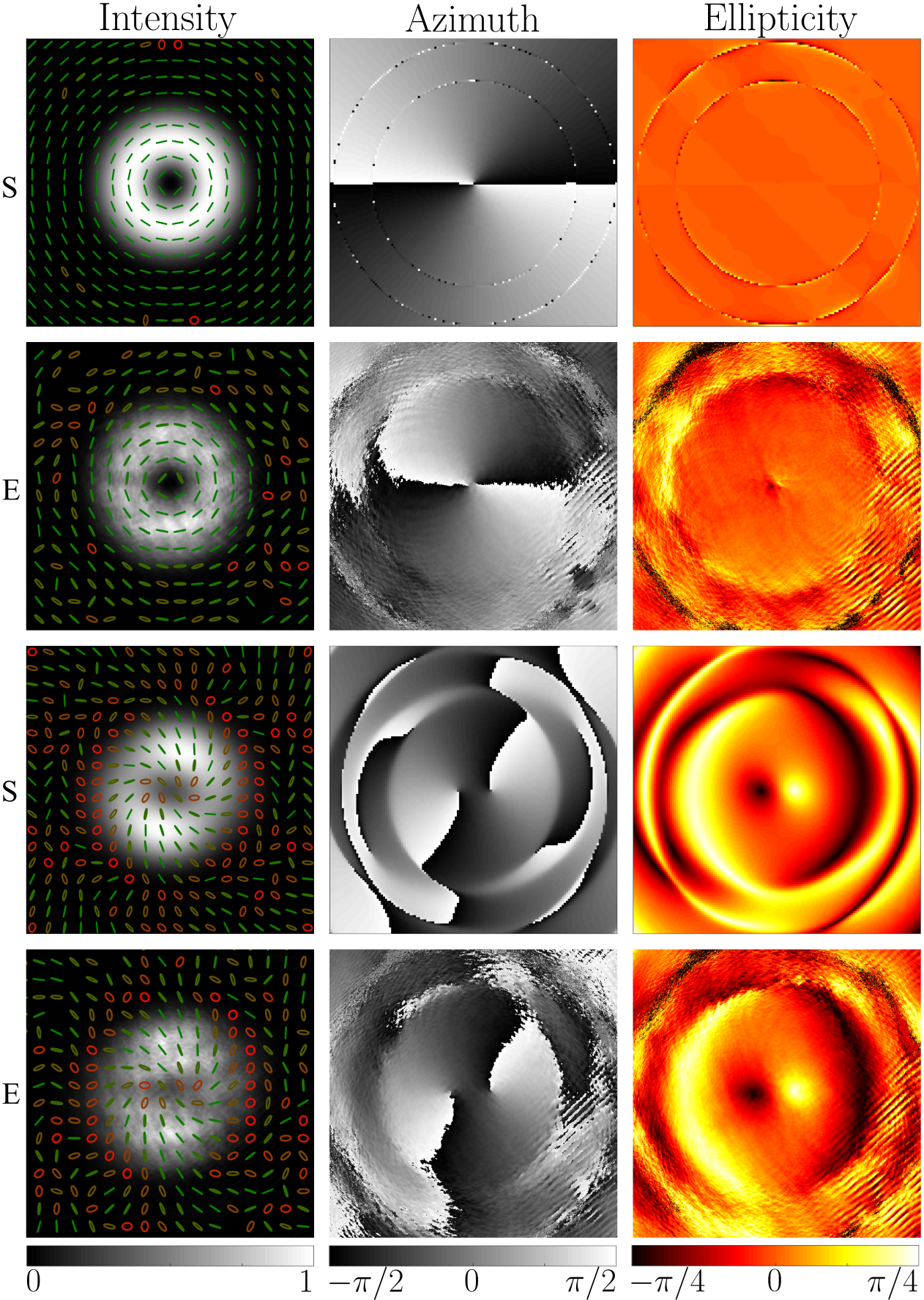}
\caption{\small Far field results for the same cases than Fig. \ref{fig:polthetaV_near}.
}
\label{fig:polthetaV}
\end{figure}

Figure \ref{fig:polthetaV_near} shows the simulated and measured intensity and polarization structure of the created beams, as well as the azimuth and ellipticity of the polarization ellipses, at the output plane of the generalized $q$-plates, for $q=1/2$ with powers $p=1$ and $p=2$. Input beam (emerging from the PSG) is linearly polarized in the vertical direction. As anticipated, the result is a vector beam with a structured linear polarization pattern in which the azimuth of the polarization vector varies as a function $2\Phi(r,\theta)$. Figure \ref{fig:polthetaV} shows the corresponding results at the Fourier plane (far field diffraction).

Polarization measurements were performed by using an imaging Stokes polarimeter, composed of a rotating quarter wave plate followed by a vertical linear polarizer. Intensity sensed at the CCD for a set of different angles (see appendix) of the wave plate gives the necessary information to calculate the Stokes parameters $S_i(x,y)$ of the measured beam. The azimuth angle gives the orientation of the polarization ellipses and is obtained as $\psi=\arctan(S_2/S_1)/2$, while the ellipticity angle gives the form of the polarization ellipses and can be calculated as $\chi=\arcsin(S_3/S_0)/2$, it ranges from  $-\pi/4$ to $\pi/4$ and it is positive for right-handed sense of rotation of the electric vector, and negative for left-handed sense of rotation \cite{goldstein}. We used these two parameters instead of the raw Stokes parameters because they show more effectively the local nature of the polarization ellipses, which helps to identify singular behaviours, as shall be seen shortly. From these parameters we plotted the polarization ellipses over the intensity distribution using a color code based on $\chi$: for $\chi = 0$ (linear polarization) we used green and for $\chi = \pm\pi/4$ (circular polarization) we used red. Intermediate colors represent elliptical polarization.

For a regular q-plate ($p=1$) the polarization field in the Fraunhofer regime shows a ``donut" intensity distribution, due to the central polarization vortex \cite{cardano}. Conversely, the beams obtained from the non-linear polynomial $q$-plates show a break in the cylindrical symmetry of the SoP and intensity distributions. In the general case it can be seen that, instead of showing a central singularity with topological charge $2q$, they show $4q$ isolated points of circular polarization (\textit{c-points}) around which polarization azimuth angle $\psi$ rotates in $\pi$. The topological charge is measured as the times $\psi$ gives a whole turn around the beams axis, so the topological charge of these c-points is $\pm 1/2$. 

\begin{figure}[H]
\centering
\includegraphics[width=0.5\columnwidth]{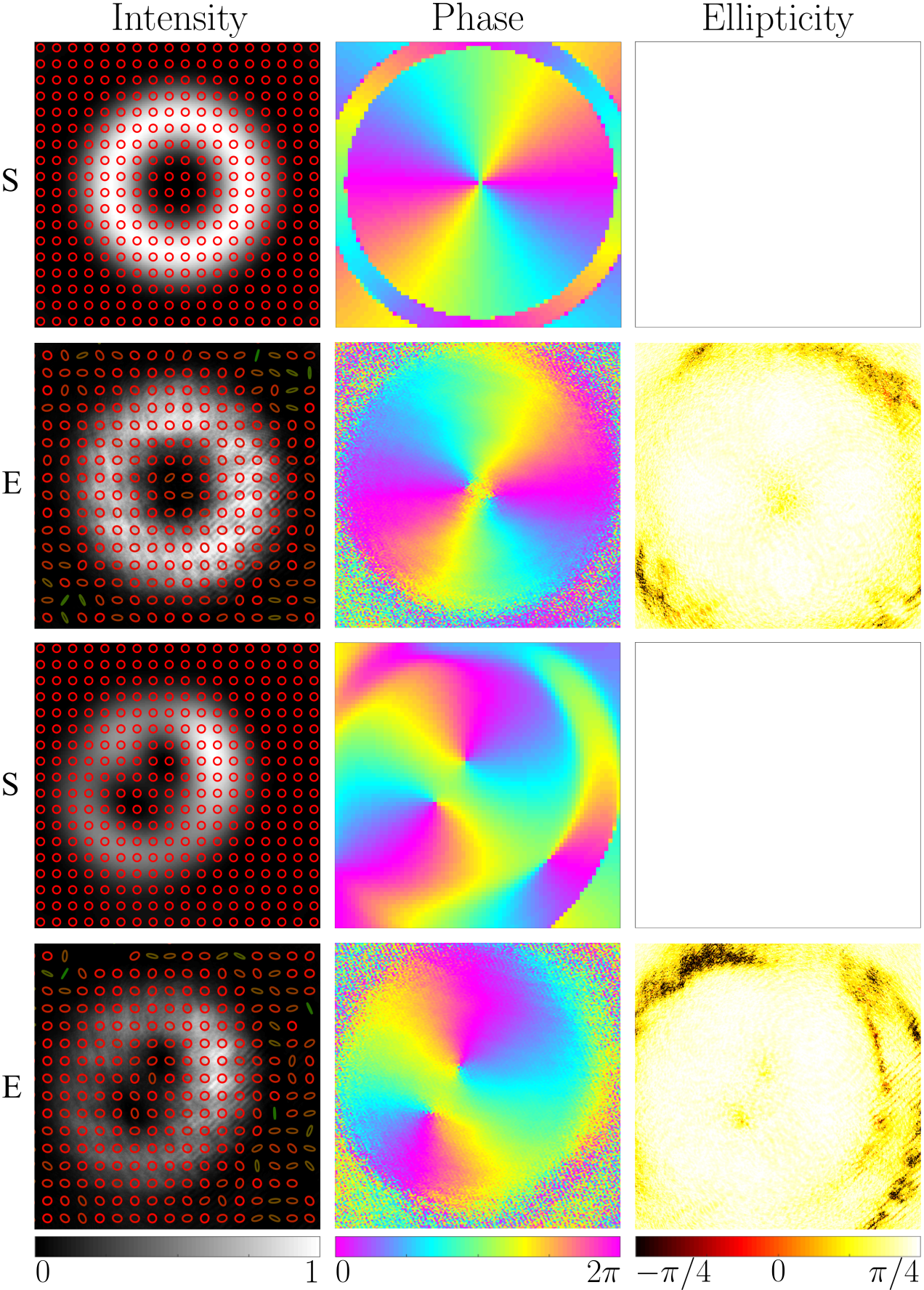}
\caption{\small Results for polynomial generalized q-plates with $q=1$, $p=1$ (first and second row) and $p=2$ (third and fourth row), in the far field regime, when incident light is left circularly polarized.
}
\label{fig:polthetaL}
\end{figure}

When input light is circularly polarized, the STOC phenomenon can be observed. Fig. \ref{fig:polthetaL} shows the intensity and polarization structure of the created beams, as well as the phase profile and ellipticity angle, at the Fourier plane (far field diffraction) of the generalized $q$-plates, for $q=1$ with powers $p=1$ and $p=2$, when the input beam is left circularly polarized.

\begin{figure}[H]
\centering
\includegraphics[width=0.5\columnwidth]{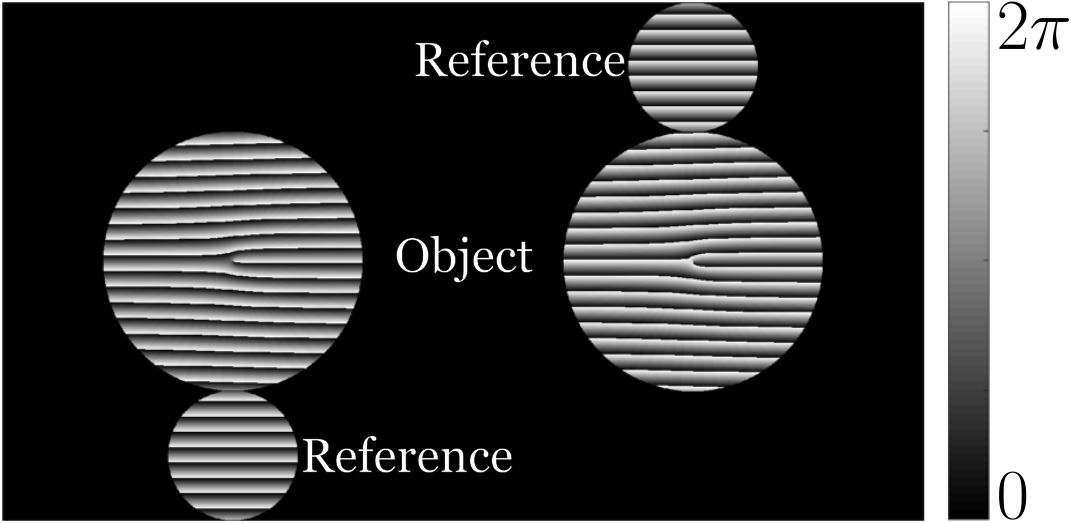}
\caption{\small Phase addressed to the SLM in order to perform phase shifting interferometry. A blazed grating is added to the $q$-plate function, obtaining a characteristic fork diagram. The reference beam phase can be shifted by displacing its respective blazed grating, which has the same period and orientation than the object grating.}
\label{fig:psi}
\end{figure}

Phase measurements were performed using a phase shifting interferometry technique. An outer region of the SLM is used to create a reference beam which interferes with the object beam at the Fourier plane. This is achieved by adding a blazed grating to the phase function of the $q$-plate and the same grating to a circular region of the SLM with uniform phase. An example of the SLM phase function addressed during a measurement with this technique is shown in Fig. \ref{fig:psi}. Displacing the reference grating results in a phase shift of the reference beam on the Fourier plane \cite{goodman}. Measuring the intensity of the interference pattern for 4 consecutive $\pi/2$ phase shifts of the reference beam, the phase of the object beam can be obtained according to
\begin{eqnarray}
    \Psi(x,y) = \arctan\left[\frac{I_4(x,y)-I_2(x,y)}{I_1(x,y)-I_3(x,y)}\right],
\end{eqnarray}
where $I_i$ represents the intensity measured in the $i$-th step \cite{creath}.

In the circularly polarized case for a linear q-plate, the phase distribution shows a central singularity (phase vortex) with topological charge $2$. Polarization remains uniformly circular, but inverts its sense of rotation to right handed, which shows the STOC phenomenon. In the non linear case, two isolated vortexes with topological charge $1$ appear, and cylindrical symmetry is lost. In the general case, when losing linearity, the central singularity of topological charge $2q$ is divided into $2q$ singularities of topological charge $\pm1$. In all these cases experimental and theoretical results show great agreement, except in the regions with very low intensity, in which measurements have a higher discrepancy.

\subsection{Polynomial $q$-plates in $r$ and $\theta$}

Following a wider scheme we implemented generalized $q$-plates of the form 
\begin{eqnarray}
    \Phi(r,\theta) =  \Phi_R(r) + \Phi_{\Theta}(\theta).
\end{eqnarray}
Thus, the phase addressed to the SLM can be thought of as two phase masks dependent of $r$ and $\theta$ respectively. We used for the sake of simplicity only polynomial functions for $\Phi_R$ and $\Phi_{\Theta}$, but the following treatment can be equally done using arbitrary 2D functions, as shall be seen in the next section.

\begin{figure}[H]
\centering
\includegraphics[width=0.5\columnwidth]{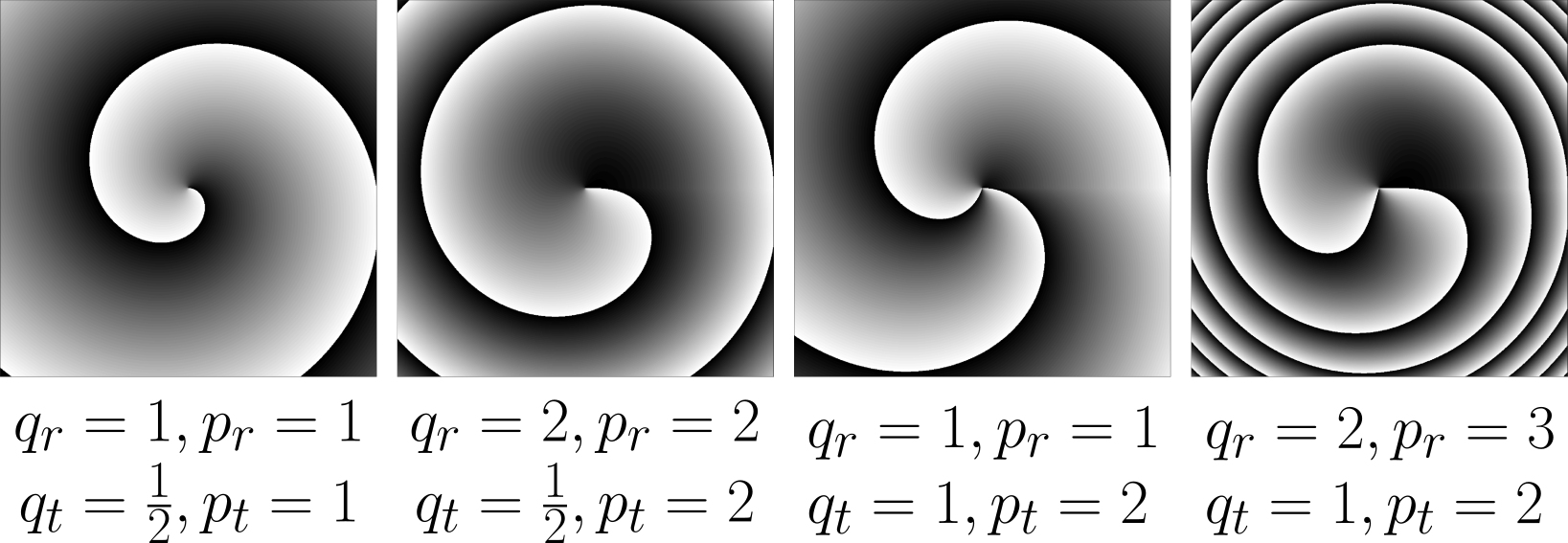}
\caption{\small Functions $2\Phi(r,\theta)$ for $q$-plates with polynomial growth both in $r$ and $\theta$.}
\label{fig:gqp}
\end{figure}

\begin{figure}[H]
\centering
\includegraphics[width=0.5\columnwidth]{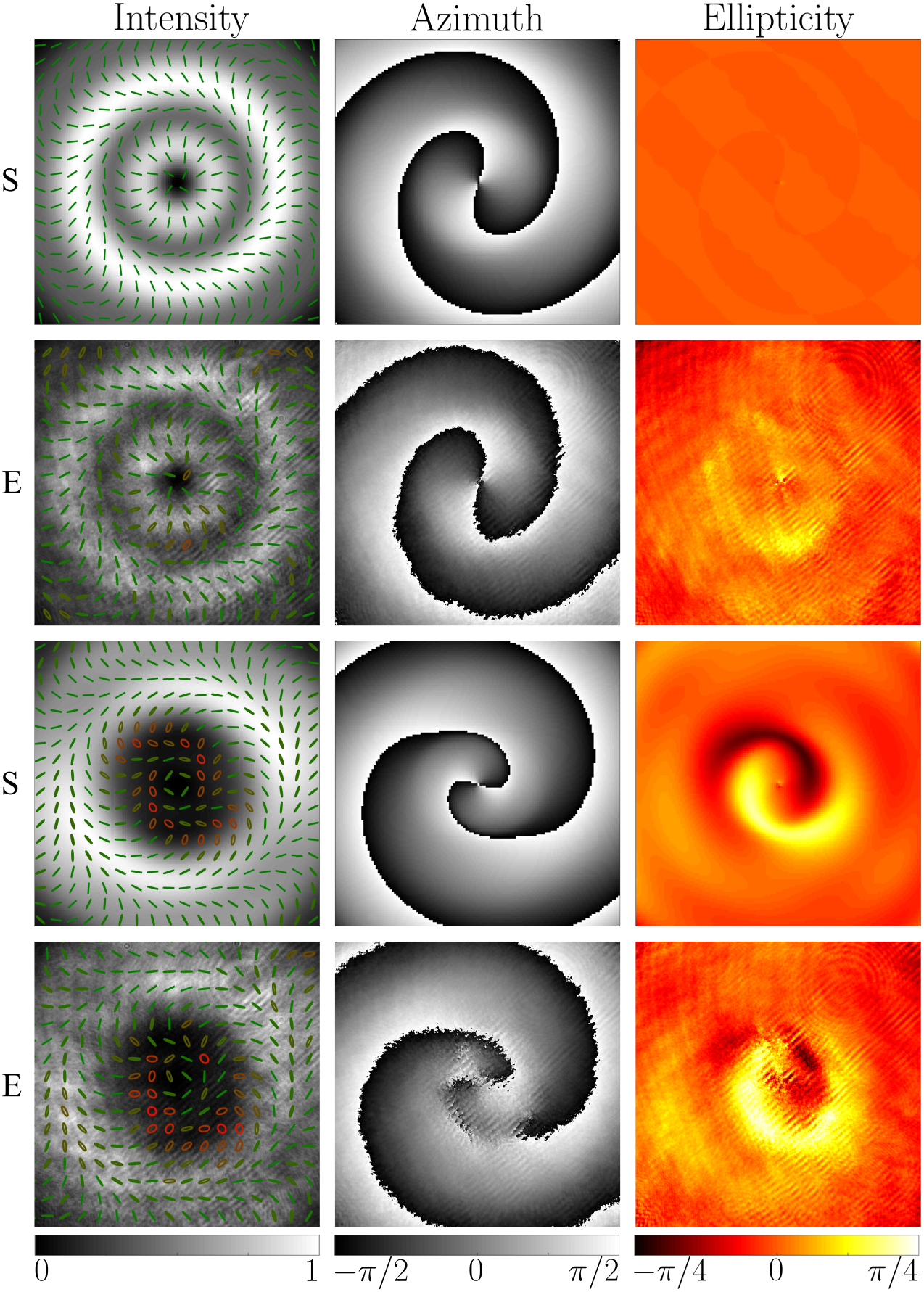}
\caption{\small Results for generalized $q$-plates with polynomial growth both in $r$ and $\theta$, in the far field regime, when input beam is vertically polarized. The first and second rows show the case with $q_r=1$, $p_r=1$, $q_t=1/2$ and $p_t=1$. The third and fourth rows show the case with $q_r=1$, $p_r=2$, $q_t=1/2$ and $p_t=2$.
}
\label{fig:CLespiralV}
\end{figure}

The generalized $q$-plate function studied is $2\Phi(r,\theta) = 2{q_r}\pi(r/r_0)^{p_r} + 2{q_t}(2\pi)^{1-{p_t}}\theta^{p_t}$. This describes a family of spiral functions like those shown in Fig. \ref{fig:gqp}. Regarding the azimuthal dependence, this function shows polynomial growth with power $p_t$ and total variation $q_t2\pi$. Radial function shows polynomial growth from the center with power $p_r$, reaching a value of $q_r\pi$ when $r = r_0$, being $r_0$ the plate radius. Results obtained in the far field, when input light is vertically polarized, are shown in Fig. \ref{fig:CLespiralV}. 

When $p_t = 1$ (linear in $\theta$), cylindrical symmetry in polarization and intensity distributions is preserved. In the case shown in the first two rows of figure \ref{fig:CLespiralV}, the result is a cylindrical vector beam with a central singularity of topological charge $2q_t = 1$. In this case, in addition, intensity and polarization azimuth varies radially. For instance, along the first concentric intensity maximum the polarization is radial, while along the next minimum it is azimuthal. This is the effect of the radial dependence.

There is a wide variety of vector beams that can be created by changing the four parameters in the expression of $\Phi$. By increasing the power $p_t$, it is observed that the central singularity splits into several singularities with lower topological charge, and the linearly polarized vector beam becomes a beam with hybrid SoP. When $q_r=1$, $p_r=2$, $q_t=\frac{1}{2}$ and $p_t=2$, as shown in the third and fourth columns of Fig. \ref{fig:CLespiralV}, there is a central intensity minimum around which polarization vector rotates, with topological charge $2q_t=1$. This singularity takes place between two regions of opposite polarization rotation handedness. The experimental results are less satisfactory in the regions with very low intensity since the relative error of the polarization measurement is increased.

The degree of freedom in $r$ allows to modulate the spacial distribution of intensity and, thus, the position of polarization singularities and critical points. We chose to show the linear and quadratic cases because they are commonly used in singular optics for representing phase functions related to axicons and lenses, respectively.

\begin{figure}[H]
\centering
\includegraphics[width=0.5\columnwidth]{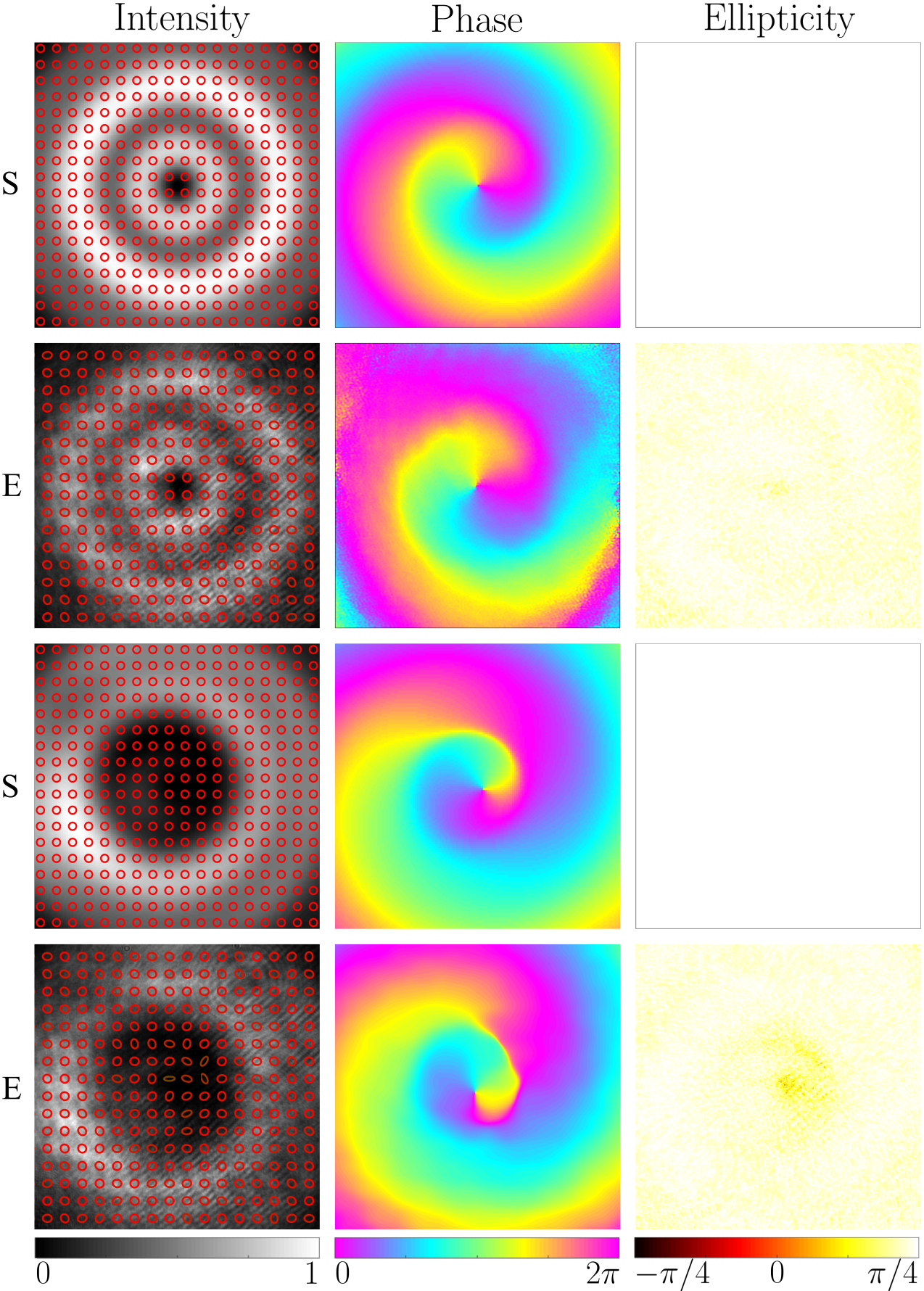}
\caption{\small Results for the same generalized $q$-plates than Fig. \ref{fig:CLespiralV}, when input beam is left circularly polarized.
}
\label{fig:CLespiralL}
\end{figure}

Figure \ref{fig:CLespiralL} shows results for the same cases than figure \ref{fig:CLespiralV}, when input light is left circularly polarized. Low intensity regions match with ellipticity minima (left handed polarization) of the case with linearly polarized input, and polarization azimuth singularities are ``replaced'' by phase singularities. A vertically polarized beam can be described as a balanced superposition of left and right circularly polarized beams, and after passing through the generalized $q$-plate, left circular polarization turns right, and vice versa. Then, it is reasonable that when input light is left circularly polarized, regions of the output beam corresponding to left circular polarization show no intensity. In this case, when growth in $\theta$ is non-linear, output intensity, phase and polarization distributions lose the cylindrical symmetry. These phase singularities can be locally seen as vortexes that carry OAM with topological charge $2q_t=\pm1$.

\section{\label{sec:multiplex}Multiplexed $q$-plates and beam superposition}

The possibility of using arbitrary functions $\Phi$ in the definition of the generalized $q$-plate and the ability of implementing them by means of SLMs, which allow pixel by pixel modulation, gives the capability of multiplexing various plates in the same device simultaneously. This creates a superposition of vector or vortex beams coming from different phase functions.

The use of a SLM leads to represent the function $\Phi$ (as a discontinuous version of itself) onto an array of square elements (pixels), which can take discrete phase values. Multiplexing can be achieved by selecting randomly two complementary sets of pixels, and representing on each set a different function. This random multiplexing scheme has been used previously, for example, to increase depth of focus of diffractive lenses \cite{iemmi2006}. This feature is possible due to the pixelated structure of the SLM and cannot be accomplished with conventional $q$-plate devices.

\begin{figure}[H]
\centering
\includegraphics[width=0.5\columnwidth]{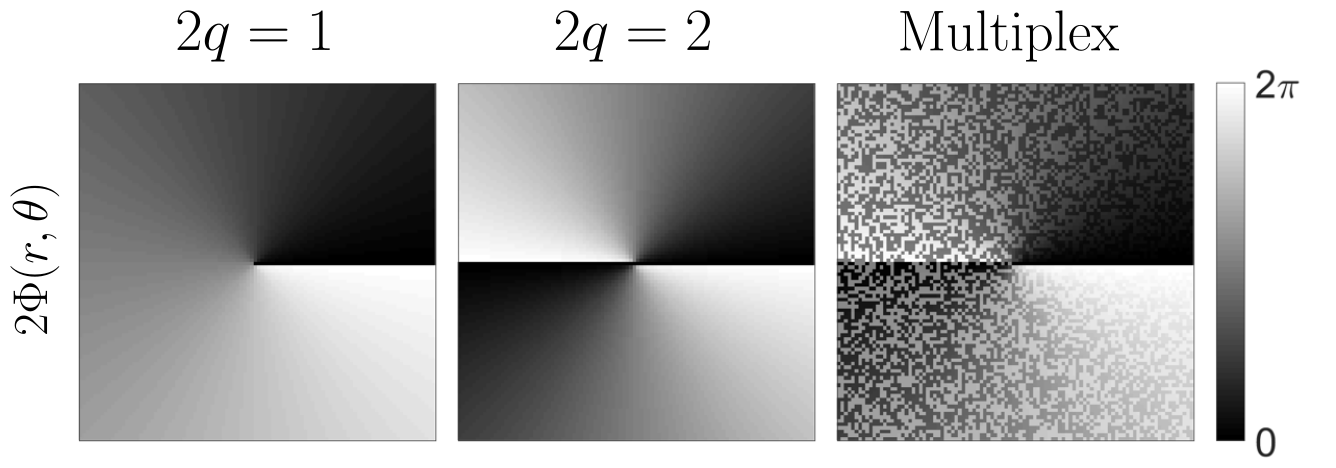}
\caption{\small $2\Phi$ functions for $q=1/2$, $q=1$ and the multiplexed $q$-plate that combines $q_1=1/2$ and $q_2=1$.}
\label{fig:multiplex}
\end{figure}

\begin{figure}[H]
\centering
\includegraphics[width=\columnwidth]{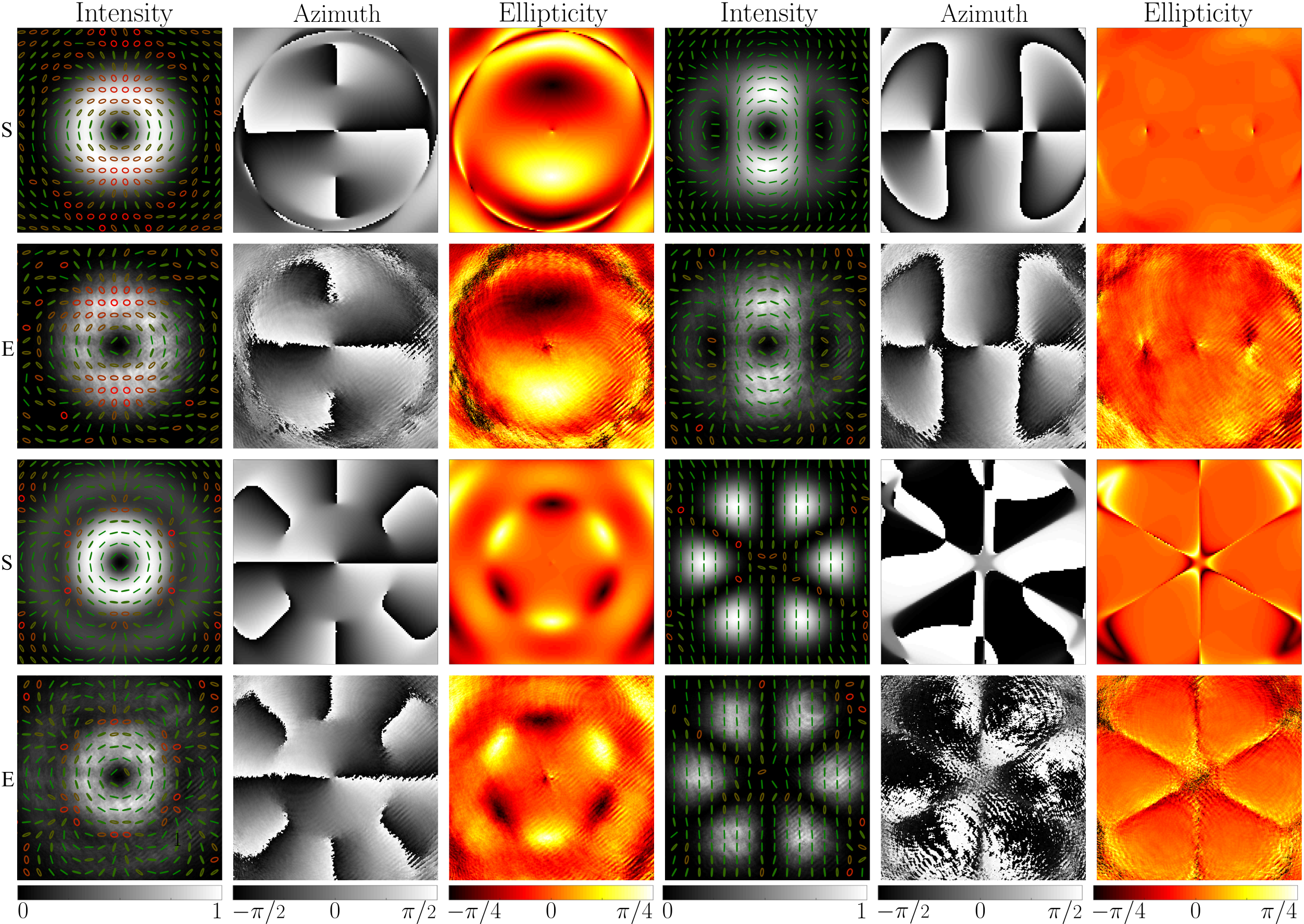}
\caption{\small Far field diffraction resulting from multiplexed $q$-plates when input beam is vertically polarized. The first and second rows show the combination of $q=1/2$ and $q=1$ (columns 1-3), and the combination of $q=1/2$ and $q=3/2$ (columns 4-6). The third and fourth rows show the combination of $q=1/2$ and $q=2$ (columns 1-3), and the combination of $q=3/2$ and $q=-3/2$ (columns 4-6).
}
\label{fig:multiV}
\end{figure}

\begin{figure}[H]
\centering
\includegraphics[width=\columnwidth]{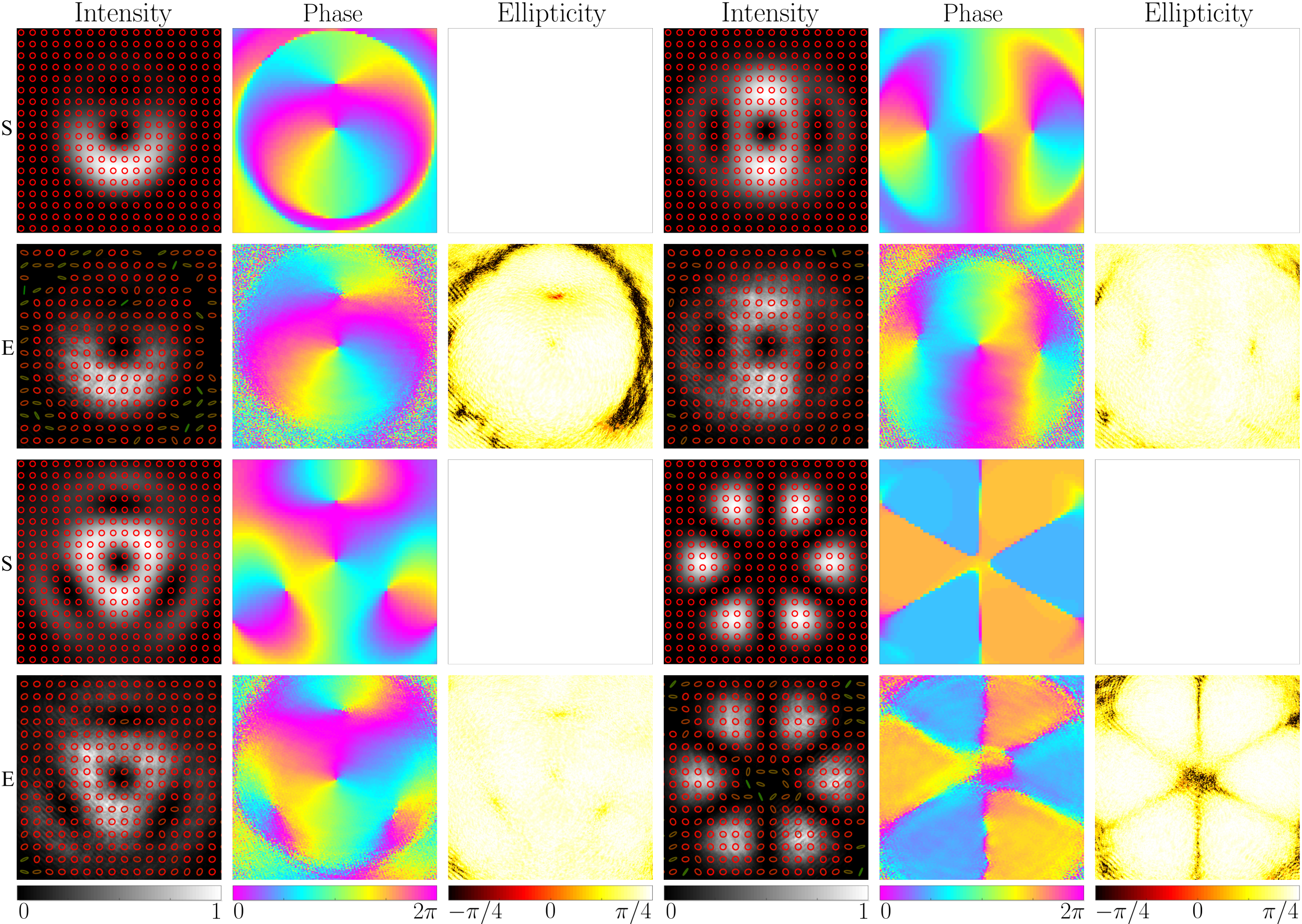}
\caption{\small Results for the same multiplexed $q$-plates than Fig. \ref{fig:multiV}, when input beam is left circularly polarized.
}
\label{fig:multiL}
\end{figure}

As a simple example, we show the result of combining pairs of linear $q$-plates, where $\Phi(\theta) = q\theta$, with different $q$ values. The functions for $q_1=1/2$ and $q_2=1$ are shown in figure \ref{fig:multiplex}, together with the discontinuous $\Phi(r,\theta)$ resultant from multiplexing both. Results for other different combinations are shown in figure \ref{fig:multiV} (for input linear polarization) and figure \ref{fig:multiL} (for input left circular polarization). The superposition of beams coming from each set of pixels is obtained. This scheme can generate superposition of alternative vector or vortex beams with arbitrary topological charges. The relative weight of the components in the superposition can be varied by changing the size of the set of pixels assigned to each component, which can be easily done given the flexibility and speed provided by the SLMs. The number of superimposed beams can be increased, and is limited by the resolution of the SLM.

Fraunhofer diffracted field of the generated beams coincides with the Fourier transform of the field at the $q$-plate plane. For a function $g(r,\theta)$ separable in polar coordinates this can be written in terms of an infinite sum of weighted Hankel transforms \cite{goodman},
\begin{eqnarray}
    \mathcal{F}\left\lbrace g(r,\theta)\right\rbrace = \sum_{k=-\infty}^{\infty} c_k (-i)^k \exp(ik\phi) \mathcal{H}_k \left\lbrace g_R (r)\right\rbrace,
\label{eq:hankel}
\end{eqnarray}
where $c_k$ is a complex coefficient and $\mathcal{H}_k$ is the Hankel transform operator of order $k$,
\begin{eqnarray}
    \mathcal{H}_k \left\lbrace g_R (r)\right\rbrace = 2\pi \int_{0}^{\infty} r g_{R} (r) J_k(2\pi r \rho) dr,
\end{eqnarray}
being $J_k$ the $k$th-order Bessel function of the first kind, and  $g(r,\theta) = g_R (r)g_{\Theta}(\theta)$.

Far field obtained from a $q$-plate with even $2q$ value only shows terms with even $k$ value in the expression of Eq. \ref{eq:hankel}, so the factor $(-i)^k = \pm 1$. When $2q$ is an odd number, only terms with odd $k$ value appear, then $(-i)^k = \pm i$. In the case of superposition of $q$-plates with topological charges of the same parity, the phase factors of both contributions differ in an even multiple of $\pm \pi/2$, i.e. they are either in phase or in counterphase. This creates uniform linearly polarized vector beams with polarization dark singularities caused by destructive interference where the added polarization vectors are parallel and have opposite phases. On the other hand, if the combined $2q$ values have different parity, the phase factors of the contributions differ in an odd multiple of $\pm \pi/2$, i.e. they are in quadrature. Hence, c-points take place at spots where added polarization vectors are orthogonal. The central singularity, characteristic to all cylindrical vortex beams irrespective of their parity, appears in all cases.

Those $q$-plates with higher $q$ value create donut shaped beams with larger radii, so in the superposition, the inner region of the beam shows the structure of the one created by the $q$-plate with lower $q$ value, while in the outer region the higher $q$ value predominates. When input polarization is left circular (Fig. \ref{fig:multiL}), only the right circular projection of the beams shown in Fig. \ref{fig:multiV} remains. Intensity minima in these cases are phase vortexes that carry OAM, as expected. If $q$-plates with opposite $q$ value are represented, the result is a beam that carries no OAM and whose polarization is uniform and with opposite handedness to the input polarization. The intensity distribution shows $4q$ azimuthal interference fringes, for which these are usually referred to as \textit{petal beams} \cite{litvin}. Azimuth angle measurement for this case shows a noisy alternation between the values $-\pi/2$ and $-\pi/2$, this error is irrelevant since both values represent indistinctly vertically polarized light. As previously addressed, regions of the images with very low intensity show less agreement with the theoretical results.

The scheme that we propose not only allows theoretically the use of generalized and multiplexed $q$-plates with arbitrary functions (of which only a few examples are given here as a demonstration), but also gives speed and flexibility to the experimental implementation, due to the use of a phase only LCoS display with high spatial resolution and video rate operation. This opens a wide range of possibilities for the creation of alternative kinds of vector and vortex beams.

\section{\label{sec:conclus} Conclusions}
We proposed a generalization of the concept of $q$-plate, including in its definition non-linear functions of both radial and azimuthal variables, which allow creating new vector or vortex beams, depending on the input state of polarization. We simulated the effect of these kind of element on uniformly vertical and left circular polarized beams and implemented them experimentally. As a result, great agreement between simulated and experimental results was achieved.

In the near field regime, function $\Phi$ determines the polarization azimuth for incident linear polarization and the phase profile for incident circular polarization. In the far field regime, it is found that when losing linearity in the azimuthal variable, the conventional central singularity of vector/vortex beams divides into several singularities of minimum topological charge. In those cases where the input light is linearly polarized, the output beam can exhibit, either c-points with topological charge $\pm\frac{1}{2}$, as well as other types of critical points of the ellipticity, or dark polarization singularities. Distribution of left and right circular polarization regions is symmetrical. Circularly polarized input beams result in the appearance of phase vortexes, which carry OAM with topological charge $\pm1$. The intensity profiles and singularity distributions depend on the particular chosen function $\Phi(r,\theta)$, which gives the chance to model distributions of any optical singularity known. This provides a tool for achieving beams with hybrid SoPs, and novel intensity distributions.

We applied this generalization to the creation of multiplexed $q$-plates, defined by discontinuous functions that consisted of different $q$-plates encoded on complementary sets of randomly picked pixels from a SLM. We obtained, in the far field regime, the superposition of vortex/vector beams generated by the individual $q$-plates involved, a result that cannot be accomplished by conventional $q$-plate devices. This has potential application in many fields including quantum and classical communications, interferometry, optical trapping and micromanipulation, and singular optics in general.

\ack{
This work was supported by UBACyT Grant No. 20020170100564BA. M.V. holds a CONICET Fellowship.}

\appendix
\section*{Appendix: Mueller-Stokes polarimetry and the effect of the beam-splitters}
\setcounter{section}{1}

The state of polarization (SoP) of a light beam (represented by its Stokes parameters) and the polarimetric behavior of a material medium (represented by its Mueller matrix), can be experimentally determined by means of a Stokes or Mueller polarimeter, respectively. A Mueller polarimeter is composed of two modules. The first one, the polarization state generator (PSG), is used to generate light beams with different and carefully selected SoPs. These beams, after interacting with the sample, are characterized by a polarization state detector (PSD). By taking an appropriate number of measurements in specific PSG and PSD configurations, a linear system of equations can be constructed, whose solution is the Mueller matrix of the sample. A Stokes polarimeter is just the PSD of a Mueller polarimeter. Given an incident light beam, its intensity must be registered for a sufficient number of properly selected PSD states. Then, the Stokes parameters can be reconstructed by solving another linear system of equations. 

In this work we used a Mueller polarimeter and also its PSD as a Stokes polarimeter. Each module was composed of a fixed linear polarizer and a fixed-retardance rotating retarder. For both instruments a measurement algorithm known as the Synchronous detection scheme was implemented. In this method each module adopts at least five measurement configurations, which correspond to equally-spaced angular positions of the fast axis of the rotating retarder. The Mueller polarimeter was calibrated by measuring the Mueller matrix of air. Since the selected Stokes polarimeter was its PSD, it became calibrated as well. For further details see reference \cite{Chironi2020}.

Since beam splitters are crucial components of the device described in Section \ref{sec:exp}, we measured their Mueller matrices in order to characterize any polarimetric properties that could affect the set up. The already described Mueller polarimeter was implemented. In the ideal case, the Mueller matrix for transmission is the identity, while the matrix for reflection it is that of an ideal mirror, such that
\begin{eqnarray}
    \mathcal{M}_{\textnormal{T}} =
    \left(
    \begin{array}{cccc}
    1 & 0 & 0 & 0 \\
    0 & 1 & 0 & 0 \\
    0 & 0 & 1 & 0 \\
    0 & 0 & 0 & 1
    \end{array}
    \right), \;
    \mathcal{M}_{\textnormal{R}} =
    \left(
    \begin{array}{cccc}
    1 & 0 & 0 & 0 \\
    0 & 1 & 0 & 0 \\
    0 & 0 & -1 & 0 \\
    0 & 0 & 0 & -1
    \end{array}
    \right).
\end{eqnarray}

The experimental Mueller matrices of both beam splitters showed some deviations from the ideal values. For instance, the Mueller matrices of the first beam splitter were
\begin{eqnarray}
    M_{\textnormal{T}} =
    \left(
    \begin{array}{cccc}
    1.00 & -0.15 & 0.00 & 0.00 \\
    -0.15 & 1.00 & 0.00 & 0.00 \\
    0.00 & 0.00 & 1.00 & -0.05 \\
    0.00 & 0.00 & -0.05 & 1.00
    \end{array}
    \right), \\
    M_{\textnormal{R}} =
    \left(
    \begin{array}{cccc}
    1.00 & 0.15 & 0.00 & 0.00 \\
    0.15 & 1.00 & 0.00 & 0.00 \\
    0.00 & -0.05 & -0.95 & 0.35 \\
    0.00 & 0.00 & -0.35 & -0.90
    \end{array}
    \right).
\end{eqnarray}

To account for these effects in our model we calculated, from these Mueller matrices, the respective Jones matrices by means of the method described in reference \cite{goldstein}. Thus,
\begin{eqnarray}
    J_{\textnormal{T}} =
    \left(
    \begin{array}{cccc}
    0.92 & 0.00 \\
    0.00 & 1.07
    \end{array}
    \right), \\
    J_{\textnormal{R}} =
    \left(
    \begin{array}{cccc}
    1.07 & 0.00 \\
    0.00 & 0.92e^{i0.36}
    \end{array}
    \right).
\end{eqnarray}
This can be interpreted as follows. During transmission, the horizontal and vertical polarization components are unequally affected, while during reflection an additional spurious phase is added to the vertical component.

Even if the amplitude modulation is unbalanced, the imbalance produced by a transmission is compensated with that produced by a subsequent reflection, and vice versa, so we can consider this as a global real factor $A$ which represents a neutral attenuator. On the other hand, by assuming that each beam-splitter introduces a different spurious phase shift, say $\alpha$ and $\beta$, the Jones matrix that describes the net effect of the SLM (equation \ref{eq:MSLM}) becomes
\begin{eqnarray}
    M_{\textnormal{SLM}} &=
    \left(
    \begin{array}{cc}
    0 & -iA\exp[i(2\Phi + \alpha)] \\
    iA\exp[i(-2\Phi + \beta)] & 0
    \end{array}
    \right) \\
    &= Ae^{i\alpha}
    \left(
    \begin{array}{cc}
    0 & -i\exp[i2\Phi] \\
    i\exp[i(-2\Phi + \beta - \alpha)] & 0
    \end{array}
    \right).
\end{eqnarray}

Therefore, the total effect of the beam-splitters can be decomposed into a complex factor $Ae^{i\alpha}$, which does not affect the polarization performance of the device, and a phase shift that can be corrected by adding a uniform term $\alpha - \beta$ to the function addressed to the first half of the LCoS. This had to be taken into account when generating the vector and vortex beams with the proposed device.

\section*{References}

\begin{thebibliography}{10}
\expandafter\ifx\csname url\endcsname\relax
  \def\url#1{{\tt #1}}\fi
\expandafter\ifx\csname urlprefix\endcsname\relax\def\urlprefix{URL }\fi
\providecommand{\eprint}[2][]{\url{#2}}

\bibitem{wang2012}
Wang J, Yang J~Y, Fazal I~M, Ahmed N, Yan Y, Huang H, Ren Y, Yue Y, Dolinar S,
  Tur M and Willner A 2012 {\em Nat. Photon.\/} {\bf 6} 488–496

\bibitem{furhapter2005}
Fürhapter S, Jesacher A, Bernet S and Ritsch-Marte M 2005 {\em Opt. Express\/}
  {\bf 13} 689--694

\bibitem{bowman2011}
Bowman R~W, Gibson G, Carberry D, Picco L, Miles M and Padgett M~J 2011 {\em
  Journal of Optics\/} {\bf 13} 044002

\bibitem{asavei2009}
Asavei T, Loke V~L~Y, Barbieri M, Nieminen T~A, Heckenberg N~R and
  Rubinsztein-Dunlop H 2009 {\em New Journal of Physics\/} {\bf 11} 093021

\bibitem{molina2004}
Molina-Terriza G, Vaziri A, Rehacek J, Hradil Z and Zeilinger A 2004 {\em
  Physical review letters\/} {\bf 92} 167903

\bibitem{mirho2013}
Mirhosseini M, Malik M, Shi Z and Boyd R~W 2013 {\em Nature Communications\/}
  {\bf 4} 2781

\bibitem{quabis}
Quabis S, Dorn R, Eberler M, Glöckl O and Leuchs G 2000 {\em Opt. Commun.\/}
  {\bf 179} 1--7

\bibitem{cheng}
Cheng W, Haus J~W and Zhan Q 2009 {\em Opt. Express\/} {\bf 17} 17829--17836

\bibitem{woerde}
Woerdemann M, Alpmann C, Esseling M and C D 2013 {\em Laser Photonics Rev.\/}
  {\bf 7} 839–854

\bibitem{gabriel}
Gabriel C, Aiello A, Zhong W, Euser T, Joly N, Banzer P, Förtsch M, Elser D,
  Andersen U, Marquardt C, Russell P and Leuchs G 2011 {\em Phys. Rev. Lett.\/}
  {\bf 106} 060502

\bibitem{marrucci2}
Marrucci L, Manzo C and Paparo D 2006 {\em Phys. Rev. Lett.\/} {\bf 96} 163905

\bibitem{Devlin896}
Devlin R~C, Ambrosio A, Rubin N~A, Mueller J~P~B and Capasso F 2017 {\em
  Science\/} {\bf 358} 896--901

\bibitem{ji}
Ji W, Lee C, Chen P, Hu W, Ming Y, Zhang L, Lin T, Chigrinov V and Lu Y 2016
  {\em Sci. Rep.\/} {\bf 6} 25528

\bibitem{holland}
Holland J~E, Moreno I, Davis J~A, Sánchez-López M~M and Cottrell D~M 2018
  {\em Appl. Opt.\/} {\bf 57} 1005--1010

\bibitem{vergara19}
Vergara M and Iemmi C 2019 {\em Phys. Rev. A\/} {\bf 100}(5) 053812

\bibitem{schulz13}
Schulz S~A, Machula T, Karimi E and Boyd R~W 2013 {\em Opt. Express\/} {\bf 21}
  16130--16141

\bibitem{tao06}
Tao S~H, Yuan X~C, Lin J and Burge R~E 2006 {\em Opt. Express\/} {\bf 14}
  535--541

\bibitem{anguita14}
{Anguita} J~A, {Herreros} J and {Djordjevic} I~B 2014 {\em IEEE Photonics
  Journal\/} {\bf 6} 1--11

\bibitem{li12}
Li X, Lan T~H, Tien C~H and Gu M 2012 {\em Nature Communications\/} {\bf 3} 998

\bibitem{lerman09}
Lerman G~M and Levy U 2009 {\em Opt. Express\/} {\bf 17} 23234--23246

\bibitem{goldstein}
Goldstein D 2003 {\em Polarized Light, Revised and Expanded\/} Optical
  engineering (CRC Press) ISBN 9780203911587

\bibitem{cardano}
Cardano F, Karimi E, Slussarenko S, Marrucci L, de~Lisio C and Santamato E 2012
  {\em Appl. Opt.\/} {\bf 51} C1--C6

\bibitem{goodman}
Goodman J~W 1996 {\em Introduction to Fourier Optics\/} (McGraw-Hill)

\bibitem{creath}
Creath K 1988 V phase-measurement interferometry techniques ({\em Progress in
  Optics\/} vol~26) ed Wolf E (Elsevier) pp 349 -- 393

\bibitem{iemmi2006}
Iemmi C, Campos J, Escalera J~C, L\'{o}pez-Coronado O, Gimeno R and Yzuel M~J
  2006 {\em Opt. Express\/} {\bf 14} 10207--10219

\bibitem{litvin}
Litvin I~A, Ngcobo S, Naidoo D, Ait-Ameur K and Forbes A 2014 {\em Opt.
  Lett.\/} {\bf 39} 704--707

\bibitem{Chironi2020}
Chironi E and Iemmi C 2020 {\em Applied optics\/} {\bf 59} 6368--6378

\end{thebibliography}
\providecommand{\newblock}{}

\end{document}